\documentclass[runningheads,fleqn]{svmult}
\usepackage{makeidx}   
\usepackage{graphicx}  
\usepackage{subeqnar}  
\usepackage{multicol}  
\usepackage{taphys}    
\makeindex             
\begin{document}
\title*{Singlet-Triplet Mixing in Superconductor-\-Ferromagnet Hybrid Devices}
\toctitle{Singlet-Triplet Mixing in Superconductor-\-Ferromagnet 
\protect\newline hybrid devices }
%
%
\titlerunning{Singlet-Triplet Mixing in SF hybrid devices}
%
\author{
M. Eschrig\inst{1}
\and 
J. Kopu\inst{1,2}
\and 
A. Konstandin\inst{1}
\and 
J.C. Cuevas\inst{1}
\and 
M.Fogelstr\"om\inst{3}
\and 
Gerd Sch\"on\inst{1,4}}
\authorrunning{Matthias Eschrig  et al.}
%
%
\institute{
Institut f\"ur Theoretische Festk\"orperphysik,
    Universit\"at Karlsruhe, 76128 Karlsruhe, Germany
\and Low Temperature Laboratory,
        Helsinki University of Technology, FIN-02015 HUT, Finland 
\and Applied Quantum Physics, MC2, Chalmers, S-41296 G\"oteborg, Sweden
\and Institut f\"ur Nanotechnologie, Forschungszentrum Karlsruhe,
    76021 Karlsruhe, Germany
     }

\maketitle              

\begin{abstract}
We develop a theory which describes
hybrid structures consisting out of superconducting and 
ferromagnetic parts. We give two examples for applications.
First, we consider a hybrid structure containing a strong
ferromagnet in the ballistic limit. Second, we study for a weak
ferromagnet the influence of a domain wall on the superconducting proximity effect.
In both cases we account quantitatively for the mixing between singlet and
triplet correlations.
\end{abstract}

\section{Introduction}
\begin{figure}
\includegraphics[width=0.33\textwidth]{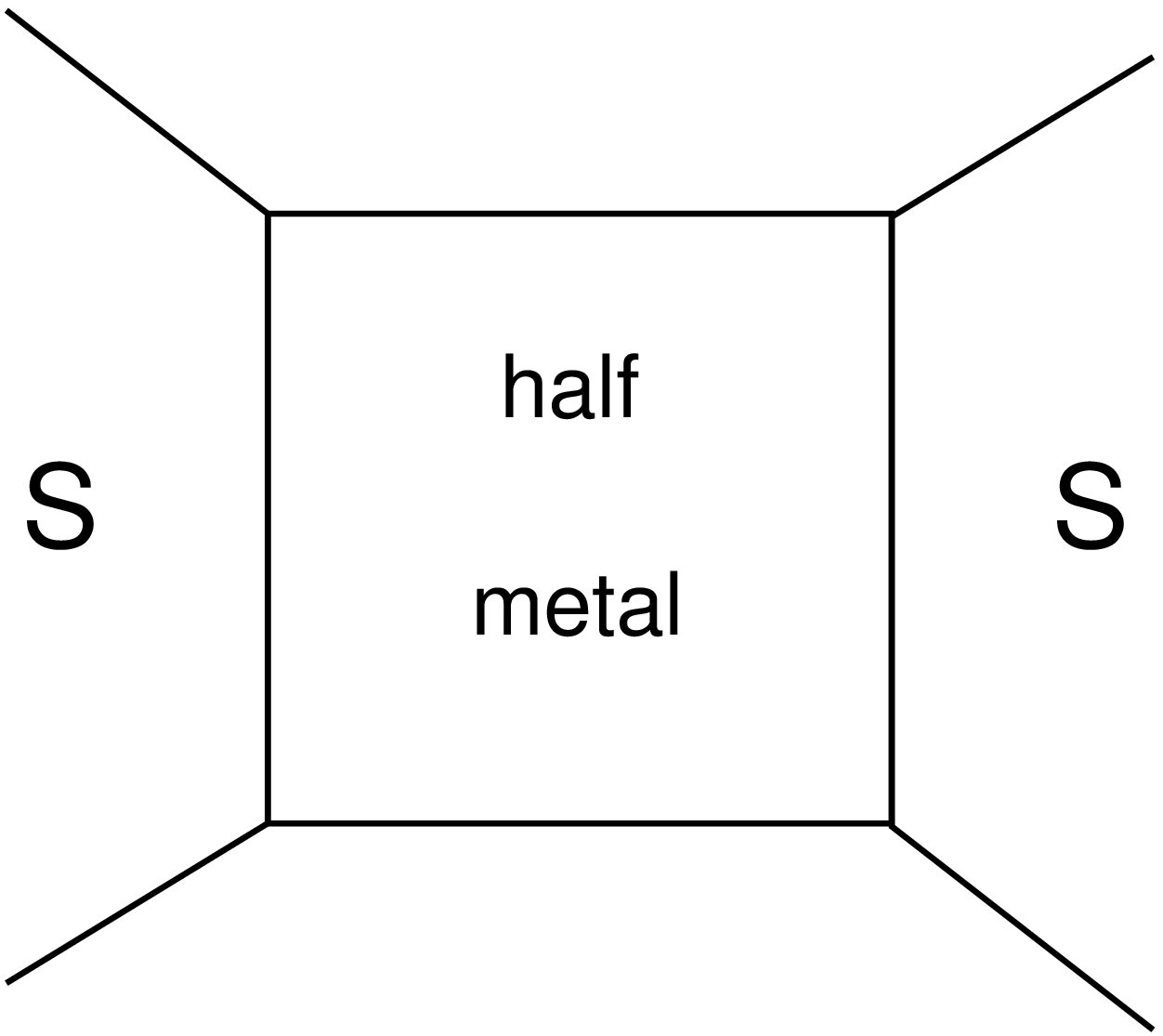}
\hfill
\includegraphics[width=0.5\textwidth]{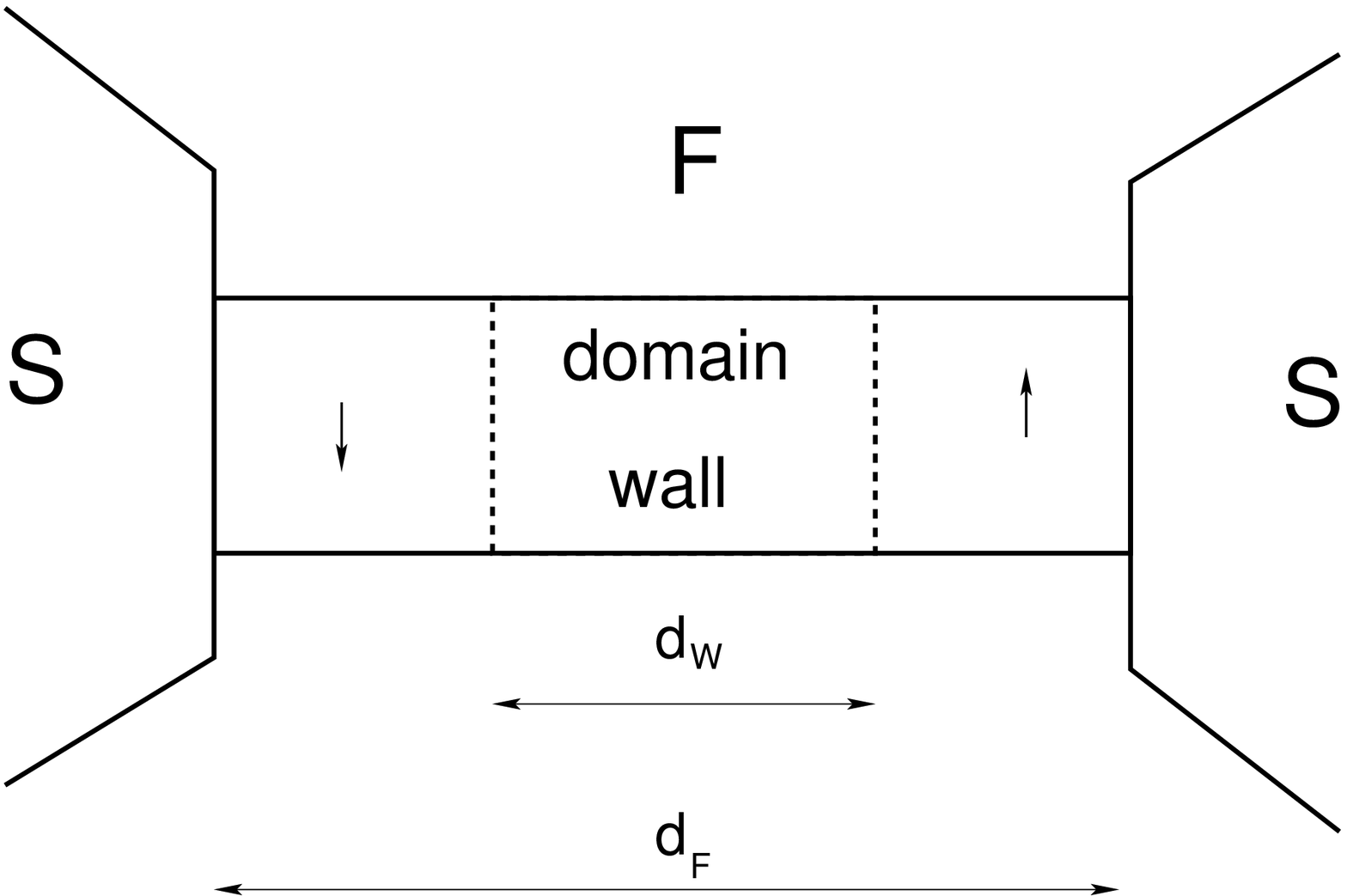}
\caption[]{Schematic picture of the studied devices. On the left, a 
ballistic heterostructure where a half metal is sandwiched between two superconductors. 
Here, the conventional proximity effect is completely suppressed as a result
of the complete spin polarization of the half metal. On the right, a diffusive weak 
ferromagnet between two superconductors. The ferromagnet contains a Bloch domain wall of
size $d_W$}
\label{eps1.1}
\end{figure}

Hybrid structures containing ferromagnetic materials became 
recently a focus of nanoelectronic research because of their relevance for
spintronics applications. Consequently, it is desirable to
understand how in the case of a superconductor
coupled to a ferromagnetic material superconducting
correlations penetrate into the ferromagnet. A powerful method
to treat such problems is the quasiclassical theory of superconductivity
developed by Larkin and Ovchinnikov and by Eilenberger \cite{Eilenberger,Larkin}.
Within this theory the quasiparticle motion is treated on 
a classical level, whereas the particle-hole  and the spin degree
of freedoms are treated quantum mechanically.
The paper consist of two parts. First we demonstrate our method for
a ballistic heterostructure containing a {\it strong} ferromagnet, for which
we chose for simplification a completely polarized ferromagnet, a half metal.
We show results for the modification of the quasiparticle density of states due to
the proximity effect in the half-metallic material in a superconductor/half metal/superconductor
device (see Fig. \ref{eps1.1}, left). 
Second, we study for a diffusive heterostructure containing a {\it weak} ferromagnet
the influence of a domain wall of the Bloch type on the quasiparticle spectrum.
The corresponding setup is shown on the right in Fig. \ref{eps1.1}.
We show results for different ratios between the domain wall width and the
superconducting coherence length.

\section{Basic Equations: Ballistic Case}
\subsection{Transport Equations}
The central quantity in quasiclassical theory of superconductivity
\cite{Eilenberger,Larkin} is 
the quasiclassical Green's function
$\check{g}({\bf p}_F,{\bf R},E,t)$ that depends on the spatial
coordinate ${\bf R}$ and time $t$. It describes quasiparticles with
energy $E$ (measured from the chemical potential) and Fermi momentum ${\bf p}_F$ 
moving along classical trajectories with direction given by the Fermi velocity
${\bf v}_F({\bf p}_F)$.\cite{Serene} 
The quasiclassical Green's function is a functional of 
self energies $\check\Sigma({\bf p}_F,{\bf R},E,t)$, which in general
include molecular fields, the superconducting order parameter 
$\Delta ({\bf p}_F,{\bf R},t)$,
impurity scattering, and external fields.
The quantum mechanical degrees of freedom of the quasiparticles show up in the
matrix structure of the quasiclassical propagator and the self energies.
It is convenient to formulate the theory using
2$\times$2 matrices in Keldysh space (denoted by a ``check'' accent), the elements of
which in turn are  4$\times$4 Nambu-Gor'kov matrices in combined
particle-hole (denoted by a ``hat'' accent) and spin space. The structure of the
propagators and self energies in Keldysh-space and particle-hole space is as follows,
\begin{equation}
\check g= \left(
\begin{array}{cc}
\hat g^R & \hat g^K \\
0 & \hat g^A
\end{array}
\right),
\quad \;
\label{gl_green3}
    \hat{g}^{R,A}=\!
    \left( \begin{array}{cc} g^{R,A} & f^{R,A} \\ \tilde{f}^{R,A} & \tilde {g}^{R,A}
      \end{array} \right)\!,~~~\hat{g}^{K}=\!
    \left( \begin{array}{cc} g^K & f^K \\ -\tilde{f}^K & -\tilde {g}^K
      \end{array} \right)\!,
\end{equation}
\begin{equation}
\label{sigma}
\check{\Sigma}=\left( \begin{array}{cc} \hat{\Sigma}^R & \hat{\Sigma}^K \\
0 & \hat{\Sigma}^A \end{array} \right),
\;
    \hat{\Sigma}^{R,A}=\!
    \left( \begin{array}{cc} \Sigma^{R,A} & \Delta^{R,A} \\ \tilde{\Delta}^{R,A} & 
     \tilde {\Sigma}^{R,A}
      \end{array} \right)\!, \hat{\Sigma}^{K}=\!
    \left( \begin{array}{cc} \Sigma^K & \Delta^K \\ -\tilde{\Delta}^K & 
      -\tilde {\Sigma}^K
      \end{array} \right)\!.
\end{equation}

The elements of the 2$\times$2 Nambu-Gor'kov matrices
are 2$\times$2 matrices in spin space, e.g. $g^R=g^R_{\alpha\beta}$ with $\alpha,\beta=\{\uparrow,
\downarrow\}$, and similarly for others.
In writing Eqs. (\ref{gl_green3}) and (\ref{sigma}) we used general symmetries, which are accounted for
by the ``tilde'' operation,
\begin{equation} 
\label{tilde}
    \tilde{X}({\bf p}_F,{\bf R},E,t)=X(-{\bf p}_F,{\bf R},-E,t)^\ast.
\end{equation}

The quasiclassical Green's functions satisfy the Eilenberger-Larkin-Ovchin\-nikov 
transport equation and normalization condition
\begin{equation}
\left[E \check \tau_3 - \check \Sigma ,
\check g
\right]_{\otimes} +
i {\bf v}_F \cdot \nabla
\check g=0,
\qquad
\label{normalize}
\check g \otimes \check g = -\pi^2 \check 1.
\label{eilen}
\end{equation}
The noncommutative product $\otimes$ combines
matrix multiplication with a convolution over the internal variables,
and $\check \tau_3=\hat \tau_3 \check 1$ is a Pauli matrix in
particle-hole space.  

The functional dependence of the quasiclassical propagator on the self energies
is given in the form of self-consistency conditions.
{\it E.g} for a weak-coupling, $s$-wave order parameter the condition reads
\begin{equation}
\hat \Delta ({\bf R},t) = \lambda \int^{E_c}_{-E_c}
\frac{dE }{4\pi i}
\langle \hat f^{K}({\bf p}_F,{\bf R},E,t) \rangle_{{\bf p}_F},
\end{equation}
where $\lambda$ is the strength of the pairing interaction, and $\langle
\hspace{2mm} \rangle_{{\bf p}_F}$ denotes averaging over the Fermi surface.
The cut-off energy $E_c$ is to be eliminated in favor of the transition
temperature in the usual manner.

When the quasiclassical Green's function has been determined,
physical quantities of interest can be calculated.
{\it E.g.} the equilibrium local density of states at position
${\bf R}$ for quasiparticles with momentum ${\bf p}_F$ and energy $E$ 
(measured from the Fermi level) reads
\begin{equation}
N ({\bf p}_F,{\bf R},E) = 
N_F \frac{1}{-4\pi i}{\rm Tr}\left[\hat \tau_3
\hat g^{R}({\bf p}_F,{\bf R},E)-\hat \tau_3 \hat g^{A}({\bf p}_F,{\bf R},E)\right].
\label{densityofstates}
\end{equation}
where $N_F$ is the density of states on the Fermi surface.

For heterostructures, the above equations must still be supplemented with
boundary conditions at the interfaces. 
As these conditions are non-trivial in quasiclassical theory, we
present these conditions in the following chapter in detail.

\subsection{Boundary Conditions }
In order to formulate the boundary conditions at an interface between two
materials, we define first an auxiliary propagator on
each side of the interface \cite{Cuevas}. 
Then we relate this auxiliary propagator to the full propagator via a
transfer matrix. This approach is completely equivalent to a full
scattering matrix approach as we will show at the end of
this paragraph. However, the repeated Andreev scattering processes
across the interface complicate the full scattering matrix approach, and
we found the current method much easier to implement and numerically
more stable \cite{Esch03,Kopu}.

We use for the auxiliary propagator the notation
$g^{\alpha,0}_{o}$ and $g^{\alpha,0}_{i}$, where the upper index 
$\alpha =\{l,r\}$ determines the side of the interface (left or right).
The lower index denotes the direction of the Fermi velocity. 
{\it Incoming} momenta (index $i$)
are those with a Fermi velocity pointing towards the interface, and {\it outgoing} momenta
(index $o$) are those with a Fermi velocity pointing away from the interface.
We formulate here the boundary conditions for clean surfaces, which conserve the
parallel component of the Fermi momentum, ${\bf p}_\parallel$.
The auxiliary propagators as function of ${\bf p}_\parallel$, energy $E$, and time $t$ are solutions
of the quasiclassical transport equation with the {\it exact} self energies, together with the normalization condition, however subject
to the {\it auxiliary boundary conditions},
\begin{eqnarray}
\check g^{\alpha,0}_{o}({\bf p}_\parallel ,E,t)&=&
\hat S^\alpha_{oi} ({\bf p}_\parallel ) \; 
\check g^{\alpha,0}_{i} ({\bf p}_\parallel ,E,t) \;
\hat S^\alpha_{io}({\bf p}_\parallel ),
\end{eqnarray}
where the following symmetries hold,
\begin{eqnarray}
\hat S^\alpha_{io}({\bf p}_\parallel )&=&\hat S^\alpha_{oi}({\bf p}_\parallel )^\dagger =
\hat S^\alpha_{oi}({\bf p}_\parallel )^{-1}. 
\end{eqnarray}
These boundary conditions are formally equivalent to boundary conditions for
an impenetrable interface, however with a surface scattering matrix $\hat S^{\alpha}_{oi}$
determined from the reflection amplitudes of the {\it full} scattering matrix as
explained below. Once these auxiliary propagators are obtained, the full propagators 
can be obtained directly, without further solving the transport equation, in the following
way. We define hopping amplitudes 
$\hat \tau^{lr}_{io}$, $\hat \tau^{rl}_{io}$, $\hat \tau^{lr}_{oi}$ and
$\hat \tau^{rl}_{oi}$, for the four channels 
(left incoming to right outgoing, right incoming to
left outgoing, left outgoing to right incoming, and right outgoing to left incoming)
which conserve the  momentum component parallel to the interface, ${\bf p}_\parallel$.
With the help of these amplitudes we solve for the {\it transfer matrices}, 
$\check t^\alpha_{i}$, for incoming trajectories from the following equations,
\begin{eqnarray}
\check t^\alpha_{i}({\bf p}_\parallel ,E,t)&=&
\hat \tau^{\alpha\beta}_{io} ({\bf p}_\parallel )\; {\check g}^{\beta,0}_{o} ({\bf p}_\parallel ,E,t)\;
\hat \tau^{\beta\alpha}_{oi}({\bf p}_\parallel ) 
\nonumber \\
&\otimes & \left( \check 1 +
{\check g}^{\alpha,0}_{i} ({\bf p}_\parallel ,E,t)\otimes \check t^\alpha_{i}({\bf p}_\parallel ,E,t) \right), 
\label{tmatrix}
\end{eqnarray}
where $(\alpha \beta)=\{(l,r),(r,l)\}$.
The corresponding
transfer matrices for outgoing trajectories are related to the ones
for incoming trajectories through the relations
\begin{eqnarray}
\check t^\alpha_{o}({\bf p}_\parallel ,E,t)&=&
\hat S^\alpha_{oi} ({\bf p}_\parallel ) \; 
\check t^\alpha_{i} ({\bf p}_\parallel ,E,t) \;
\hat S^\alpha_{io}({\bf p}_\parallel ).
\end{eqnarray}
Particle conservation requires certain symmetries between the hopping elements, which are,
\begin{eqnarray}
\hat \tau^{\beta\alpha}_{oi}({\bf p}_\parallel )&=&\hat \tau^{\alpha\beta}_{io}({\bf p}_\parallel )^\dagger=
\hat S^\beta_{oi}({\bf p}_\parallel )\hat \tau^{\beta\alpha}_{io}({\bf p}_\parallel )\hat S^\alpha_{oi}({\bf p}_\parallel ).
\end{eqnarray}
Consequently, only one amplitude, e.g. $\hat \tau^{lr}_{io}$, contains free material parameters, 
the other three amplitudes depend on it. 
The hopping amplitudes are defined via the transmission
amplitudes of the full scattering matrix as is shown below. 
Formally, they describe
the modifications of the decoupled problem due to virtual hopping processes to the opposite side.

The {\it full propagators}, fulfilling
the desired boundary conditions at the interface, can now be easily calculated.
For incoming trajectories they are obtained from
\begin{eqnarray}
&&\check g^\alpha_{i}({\bf p}_\parallel ,E,t)\!\!=\!\! 
\check g^{\alpha,0}_{i} ({\bf p}_\parallel ,E,t)\nonumber \\ 
&+& \left( \check g^{\alpha,0}_{i}({\bf p}_\parallel ,E,t) + i\pi\check 1 \right) \otimes
\check t^\alpha_{i} ({\bf p}_\parallel ,E,t)
\otimes \left(\check g^{\alpha,0}_{i}({\bf p}_\parallel ,E,t) - i\pi\check 1\right),
\label{eq7}
\end{eqnarray}
and for outgoing trajectories from
\begin{eqnarray}
&&\check g^\alpha_{o}({\bf p}_\parallel ,E,t)\!\!=\!\!
\check g^{\alpha,0}_{o} ({\bf p}_\parallel ,E,t)\nonumber \\ 
&+&
\left(\check g^{\alpha,0}_{o}({\bf p}_\parallel ,E,t) - i\pi\check 1\right) \otimes
 \check t^\alpha_{o} ({\bf p}_\parallel ,E,t)
 \otimes \left(\check g^{\alpha,0}_{o}({\bf p}_\parallel ,E,t) + i\pi\check 1\right).
\label{eq8}
\end{eqnarray}
In this formulation, the boundary problem effectively reduces to
calculating the auxiliary Green's functions for perfectly reflecting
interfaces. 
Numerically this is an extremely simple task, {\it e.g.},
employing the procedure of Riccati parameterization as explained below. 
Afterwards the
boundary Green's functions for the partially transmitting interface
can be obtained directly from Eqs.~(\ref{eq7}) and (\ref{eq8}), 
since solving for the necessary transfer matrices (\ref{tmatrix}) 
only involves a matrix inversion.

In the transfer-matrix description, the phenomenological
parameters containing the microscopic information of the interface are
the two surface scattering matrices $S^{l,r}_{oi}$ and the 
hopping amplitude $\tau^{lr}_{io}$. All three quantities are 2x2 matrices
in spin space. As $S^{\alpha}_{io}$ is unitary, it depends apart from the
direction of the quantization axis on
two parameters, a scalar scattering phase and a spin mixing angle
(or, equivalently, a spin rotation angle).
Similarly, as $\tau^{lr}_{io}$ is hermitian, it also depends on two parameters,
one describing spin conserving transmission and the other spin flip 
transmission. 
All remaining quantities defined above are related to these material parameters
by symmetries.

The particle-hole structures of the surface scattering matrix and the
hopping amplitude are given by,
\begin{equation}
\hat S^{\alpha}_{oi}({\bf p}_\parallel)=\left(
\begin{array}{cc}
S^{\alpha}_{oi}({\bf p}_\parallel) & 0 \\
0 & \tilde S^{\alpha}_{oi}({\bf p}_\parallel)
\end{array}
\right), \; \;
\hat \tau^{\alpha,\beta}_{io}({\bf p}_\parallel)=\left(
\begin{array}{cc}
\tau^{\alpha \beta}_{io}({\bf p}_\parallel) & 0 \\
0 & 
\tilde \tau^{\alpha\beta}_{io}({\bf p}_\parallel) 
\end{array}
\right),
\end{equation}
with the hole components,
\begin{eqnarray}
\tilde S^{\alpha}_{oi} ({\bf p}_\parallel)&=&
S^{\alpha}_{io}(-{\bf p}_\parallel)^\ast=
S^{\alpha}_{oi}(-{\bf p}_\parallel)^{tr} \\
\tilde \tau^{\alpha\beta}_{io} ({\bf p}_\parallel)&=& \tau^{\alpha\beta}_{oi} ({-\bf p}_\parallel)^\ast 
= 
S^{\alpha}_{oi}(-{\bf p}_\parallel)^\ast \; \tau^{\alpha\beta}_{io}(-{\bf p}_\parallel)^\ast \; S^{\beta}_{oi}(-{\bf p}_\parallel)^\ast
.
\end{eqnarray}

Finally, we relate the parameters $S^{\alpha}_{oi}$ and $\tau^{\alpha\beta}_{io}$ to the 
{\it full normal state scattering matrix} $\hat {\bf S}$, 
\begin{eqnarray}
\label{scatt}
&&\hat {\bf S}=
\left(
\begin{array}{cc}
\hat {\bf S}_{ll} & \hat {\bf S}_{lr} \\ \hat {\bf S}_{rl} & -\hat {\bf S}_{rr}
\end{array}
\right).
\end{eqnarray}
The scattering matrix is diagonal in particle-hole space, with diagonal
components
\begin{eqnarray}
\hat {\bf S}_{\alpha\alpha}&=&
(1+\pi^2\hat\tau^{\alpha\beta}_{oi} \hat\tau^{\beta\alpha}_{io})^{-1} \; (1-\pi^2\hat\tau^{\alpha\beta}_{oi} \hat\tau^{\beta\alpha}_{io})\;
\hat S^\alpha_{oi} ,
\end{eqnarray}
and off-diagonal components
\begin{eqnarray}
\hat {\bf S}_{\alpha\beta}&=&
(1+\pi^2\hat\tau^{\alpha\beta}_{oi} \hat\tau^{\beta\alpha}_{io})^{-1} \; 2\pi \hat\tau^{\alpha\beta}_{oi} .
\end{eqnarray}
These identities serve as a precise definition
of the auxilary parameters of the theory, 
$S^l_{oi}({\bf p}_\parallel)$, $S^r_{oi}({\bf p}_\parallel)$, and $\tau^{lr}_{io}({\bf p}_\parallel)$
in terms of the
physical parameters of the full scattering matrix.

\subsection{Riccati parameterization}

The method of the Riccati parameterization \cite{Nagai,Schopohl,Eschrig} of
the quasiclassical Green's functions has proved powerful during recent years.
It accounts automatically for the normalization condition. A corresponding parameterization
for distribution functions was applied to the Keldysh part of the Green's functions.
The combined equations in their general form are \cite{Eschrig}
\begin{eqnarray} 
\label{ricc2}
    \hat{g}^K\!&=&-2i\pi \hat{N}^R 
	\otimes \!\!\left(\!\begin{array}{cc}
     (x-\gamma^R\otimes\tilde{x}\otimes\tilde{\gamma}^A) 
	& -(\gamma^R\otimes\tilde{x}-x\otimes\gamma^A)
    \\ -(\tilde{\gamma}^R\otimes x-\tilde{x}\otimes\tilde{\gamma}^A)
        & (\tilde{x}-\tilde{\gamma}^R\otimes x \otimes \gamma^A)
	\end{array}\!\right) 
	\otimes \hat{N}^A,
	\nonumber
\end{eqnarray}
\begin{eqnarray} 
\label{ricc1}
    \hat{g}^{R,A}\!&=&\pm i\pi \hat{N}^{R,A} 
	\otimes \!\!\left(\!\begin{array}{cc}
     (1+\gamma^{R,A} \otimes \tilde{\gamma}^{R,A}) & 2\gamma^{R,A}
    \\ -2\tilde{\gamma}^{R,A} &
    -(1+\tilde{\gamma}^{R,A} \otimes \gamma^{R,A})
	\end{array}\!\right), 
\end{eqnarray}
with 
\begin{equation}
\label{ricc3}
     \hat{N}^{R,A}=\!\!
	\left(\!\!\begin{array}{cc}
     (1-\gamma^{R,A}\otimes\tilde{\gamma}^{R,A})^{-1} & 0 
    \\ 0 &
      (1-\tilde{\gamma}^{R,A}\otimes\gamma^{R,A})^{-1}
	\end{array}\!\!\right). 
\end{equation}
Thus, the problem is reduced to the solution for 
two 2$\times$2 matrices in spin space,
$\gamma^R$ and $x$. Using fundamental symmetries
(particle--hole, retarded--advanced) like Eq.~(\ref{tilde}) and 
\begin{equation}
\gamma^A({\bf p}_F,{\bf R},E,t)=\gamma^R(-{\bf p}_F,{\bf R},-E,t)^{tr},
\end{equation}
where $tr$ denotes the spin-matrix transpose operation, the full retarded, advanced and Keldysh
Green's functions are obtained from $\gamma^R$ and $x$.
The transport equations for the functions $\gamma^R({\bf p}_F,{\bf R},E,t)$ and $x({\bf p}_F,{\bf R},E,t)$
are
\begin{eqnarray}
\label{ricc1a}
2E\gamma^R+
i {\bf v}_F \nabla \gamma^R
&=&\gamma^R\otimes\tilde{\Delta}^R\otimes\gamma^R+\Sigma^R\otimes
\gamma^R-\gamma^R\otimes\tilde{\Sigma}^R-\Delta^R,
\end{eqnarray}
and
\begin{eqnarray}
\label{ricc2a}
i\partial_t x
+ i {\bf v}_F \nabla x 
&+&(-\gamma^R\otimes\tilde{\Delta}^R-\Sigma^R)\otimes x 
+x \otimes (-\Delta^A\otimes\tilde{\gamma}^A+\Sigma^A)
\nonumber \\
&=& -\gamma^R\otimes\tilde{\Sigma}^K\otimes\tilde{\gamma}^A+\Delta^K\otimes
\tilde{\gamma}^A+\gamma^R\otimes\tilde{\Delta}^K-\Sigma^K. 
\end{eqnarray}

\section{Results: Ballistic Case}

As an example we consider here the Andreev quasiparticle spectrum in a
half-metallic ferromagnet between two singlet superconductors.
This corresponds to the device shown on the left side in Fig. \ref{eps1.1}.
In this case, on the half-metallic side only quasiparticles with one
spin direction with respect to the quantization axis are itinerant.
The stable equilibrium configuration
of such a system is a $\pi $-junction, in which the phase of the singlet order parameter
on both sides of the interface differs by $\pi$ \cite{Esch03}.
Characteristic of such a heterostructure is the presence of triplet correlations near the interface.
\begin{figure}
\sidecaption
\includegraphics[width=0.6\textwidth]{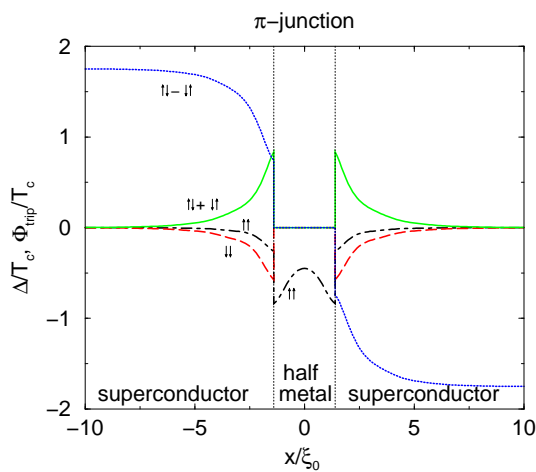}
\caption[]{Self consistent order parameter and triplet correlations in an
S/HM/S $\pi$-junction. The calculations are for temperature $T=0.05T_c$,
and for $\tau_{\downarrow , \uparrow}/\tau_{\uparrow , \uparrow}=0.7$}
\label{eps2.1}
\end{figure}
Shown in Fig. \ref{eps2.1} are the singlet order parameter and all three triplet components which
are induced near the interface by the spin rotation effect \cite{Tokuyasu88}.
The spin rotation effect is taken into account via a surface scattering
matrix $\hat S_{oi}= \exp(i\theta \sigma_z/2) \hat 1 $ 
at the superconducting side of the interface,
where $\theta $ defines a spin-rotation angle 
and $\sigma_z$ denotes the Pauli spin matrix \cite{Tokuyasu88,Fogelstrom00}. 
Generally, the value of $\theta $ depends on the angle of impact, 
$\psi $ \cite{Tokuyasu88}
and can approach values of the order of $\pi $ for strong band 
splitting \cite{Barash02}. For definiteness, we present results for
$\theta = 0.75 \pi \; \cos \psi $.
On the half-metallic side the scattering matrix has no spin structure.
Further, we use the hopping amplitude $\tau_{io} = (1+S_{io})
\tau_0 \cos \psi $, where
$\tau_0 =(\tau_{\uparrow , \uparrow},\tau_{\downarrow , \uparrow})^{\rm T}
$ 
is determined by the two spin scattering channels
from the superconductor to the half-metallic spin-up band.
This reflects a spin rotation during transmission which is half of the spin
rotation during reflection.
The $\cos \psi$ factor accounts for the reduced transmission 
at large impact angles.
We present results for $\tau_{\downarrow , \uparrow}/ \tau_{\uparrow , \uparrow}=
0.7$ and $0.1$, $2\pi \tau_{\uparrow , \uparrow}=1.0$.
The length of the half metallic region is $3 \xi_0$ 
with the coherence length $\xi_0= v_f/2\pi T_c$.  We assume
cylindrical Fermi surfaces.

\begin{figure}
\includegraphics[width=1.0\textwidth]{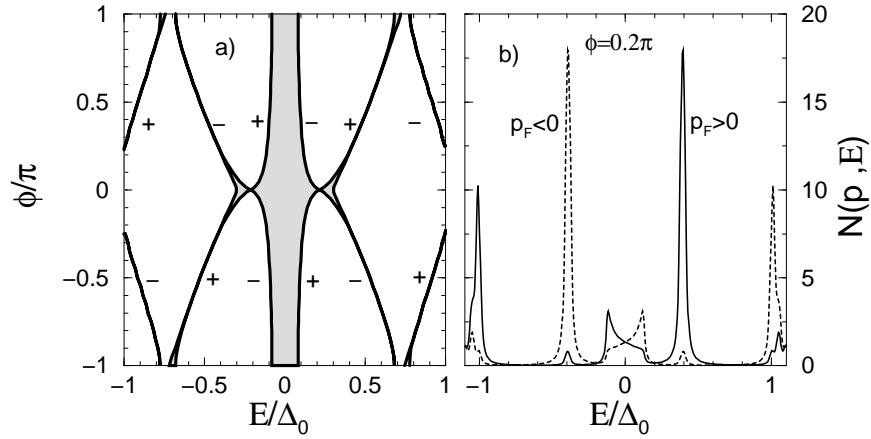}
\caption[]{Density of states at $T=0.05T_c$ for quasiparticles with normal
impact at the half-metallic side of the left interface, for
$\tau_{\downarrow , \uparrow}/\tau_{\uparrow , \uparrow}=0.7$.
a) Dispersion of the maxima of the density of states as function of
phase difference. Gapped regions are white, ungapped regions gray. 
The signs indicate the direction of the current carried by the
Andreev states.
b) As an example we show spectra for a selected phase difference }
\label{eps3.1}
\end{figure}

\begin{figure}
\includegraphics[width=1.0\textwidth]{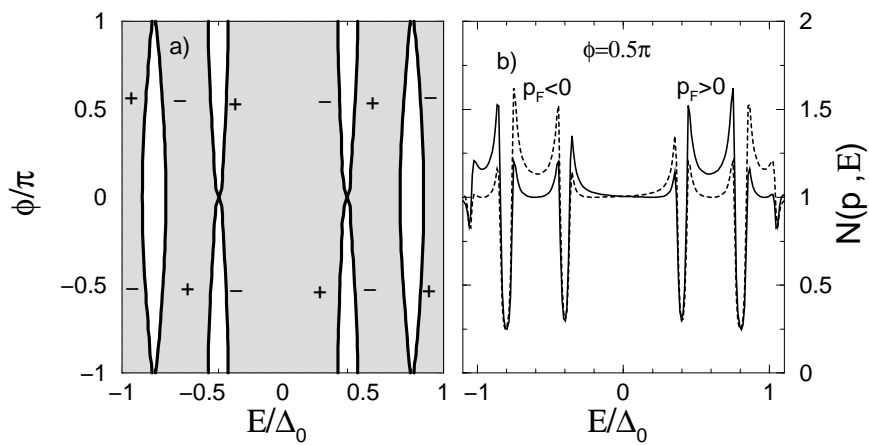}
\caption[]{The same as Fig. \ref{eps3.1} for
$\tau_{\downarrow , \uparrow}/\tau_{\uparrow , \uparrow}=0.1$}
\label{eps4.1}
\end{figure}

We have studied the behavior of such a heterostructure as a Josephson device. If a phase difference
$\phi $ in the order parameter is present between the left and right end of the heterostructure,
the quasiparticle spectrum in the half metal is modified. In Figs. \ref{eps3.1} and \ref{eps4.1}
we show results of our calculations for two sets of parameters.


The spectra consist out of Andreev quasiparticle bands, separated by gaps. 
In Figs. \ref{eps3.1} a) and \ref{eps4.1} a)
we show these gapped-regions (white) and the non-gapped regions
(grey) of the density of states at the half-metallic side of the left interface
as a function of the phase difference.
One characteristic spectrum for fixed phase difference for 
both positive and negative momentum direction is shown in 
Figs. \ref{eps3.1} b) and \ref{eps4.1} b).
The dispersion of the maxima in the density of states with phase difference
determines the sign of the Josephson current trough the device.
We indicate in Figs. \ref{eps3.1}a) and
\ref{eps4.1}a) by $+$ and $-$ Andreev states, which carry current 
in positive and negative direction respectively.
It can be seen that the sign of the Josephson current corresponds to
the sign of $d E(\phi )/d\phi $, where $E(\phi)$ denotes the dispersion
of the maxima in the density of states. 
This is analogous to the current carried by a bound state $E_b$ dispersing with phase $\phi$,
which is given by $J_b\sim (d E_b /d\phi ) n(E_b)$, where $n(E_b) $ is the equilibrium fermion 
distribution function.

\section{Basic Equations: Diffusive Case}
\subsection{Transport Equations}
The fundamental quantity for diffusive transport is the Usadel Green's function \cite{Usadel}, which
is the momentum average of the quasiclassical Green's function
$\check g({\bf R},E,t)=\langle \check g({\bf p}_F,{\bf R},E,t) \rangle_{{\bf p}_F}$. 
It is a functional of momentum averaged self energies 
$\check \Sigma({\bf R},E,t)=\langle \check \Sigma({\bf p}_F,{\bf R},E,t) \rangle_{{\bf p}_F}$.
The structures of $\check g$ and $\check \Sigma$ are the same as in
Eqs. (\ref{gl_green3}) and (\ref{sigma}). Eq. (\ref{tilde}) is replaced by
\begin{equation} 
\label{ustilde}
    \tilde{X}({\bf R},E,t)=X({\bf R},-E,t)^\ast.
\end{equation}
The Usadel Green's function obeys the following
transport equation and normalization condition \cite{Usadel},
\begin{equation} 
\label{gl_usdl}
    \left[ E \hat{\tau}_3 \check{1} -\check{\Sigma}\, , \, 
    \check{g} \right]_\otimes +\frac{D}{\pi} \nabla_j \left( \check{g} 
	\otimes \nabla_j \check{g}
    \right) = 0, \qquad
\check g \otimes \check g = -\pi^2 \check 1.
\end{equation} 
Here, summation over the repeated index $j$ is implied. $\check{\Sigma}$ does not contain the non-magnetic
impurity scattering self energy anymore.
We will solve these equations by using the Riccati parameterization technique and present below
the resulting equations for the diffusive case.
The corresponding equations in a finite vector potential are obtained by applying a
gauge transformation to Eqs. (\ref{gl_usdl}) (and to Eqs. (\ref{usricc1}), (\ref{usricc2}) below).

\subsection{Boundary Conditions }
In the diffusive limit the boundary conditions are formulated in terms of the
momentum averaged transfer matrices and Green's functions. The resulting boundary conditions 
for spin-conserving interfaces were
found by Nazarov \cite{naz} and follow from the transfer-matrix approach outlined above \cite{Kopu}. 
They are 
\begin{eqnarray}
\label{gl_naz}
\sigma_\beta \check{g}^\beta \frac{d}{dz}\check{g}^\beta
& = &\frac{1}{S R_b}~\frac{2\pi^2 T[\check{g}^\beta \, ,\, \check{g}^\alpha ]}
{4\pi^2-T \left( \{\check{g}^\beta \, ,\, \check{g}^\alpha \}+2\pi^2 \right)} \nonumber \\
\sigma_\beta \check{g}^\beta \frac{d}{dz}\check{g}^\beta
& = &
\sigma_\alpha \check{g}^\alpha \frac{d}{dz}\check{g}^\alpha  ,
\end{eqnarray}
where $z$ is the coordinate along the interface normal, 
$\sigma_{\alpha (\beta )}$ and $\check{g}^{\alpha (\beta )}$ refers to the conductivity
and the Keldysh Green's function on side $\alpha (\beta )$ of the interface,
$S$ is the surface area of the contact, $R_b$ the boundary resistance
and $T$ the transmission probability,
$
T=\frac{4\pi^2 |\tau |^2}{(1+\pi^2 |\tau |^2)^2}.
$

\subsection{Riccati parameterization}
We use the same parameterization, Eqs.~(\ref{ricc1})--(\ref{ricc3}),
for the momentum averaged Green's functions. Note that the such defined diffusive
quantities $\gamma^R({\bf R},E,t)$ and $x({\bf R},E,t)$ are by no means in a simple way related to
the ballistic quantities $\gamma^R({\bf p}_F,{\bf R},E,t)$ and $x({\bf p}_F,{\bf R},E,t)$.
Consequently, also the structure of the transport equations is very different. In particular
the gradient terms in the transport equations for these quantities
are modified with respect to the gradient terms in the transport equations for the ballistic case.
Instead of Eqs. (\ref{ricc1a}) and (\ref{ricc2a}) we obtain for the diffusive case
\begin{eqnarray}
\label{usricc1}
2E\gamma^R
&-&iD \bigg[ \nabla^2\gamma^R
+(\nabla_j\gamma^R)\otimes
\frac{\tilde{f}^R}{i\pi}\otimes(\nabla_j\gamma^R)  \bigg]
\nonumber
\\
&=&\gamma^R\otimes\tilde{\Delta}^R\otimes\gamma^R+\Sigma^R\otimes
\gamma^R-\gamma^R\otimes\tilde{\Sigma}^R-\Delta^R, 
\end{eqnarray}
and
\begin{eqnarray}
\label{usricc2}
\nonumber
i\partial_t x
&-&iD \bigg[
\nabla^2 x
-(\nabla_j\gamma^R)\otimes
\frac{\tilde{g}^K}{i\pi}\otimes(\nabla_j\tilde{\gamma}^A) \qquad \qquad \\ 
&+&(\nabla_j\gamma^R)\otimes
\frac{\tilde{f}^R}{i\pi}\otimes(\nabla_j x) 
+(\nabla_j x)\otimes
\frac{{f}^A}{i\pi}\otimes(\nabla_j\tilde{\gamma}^A) \bigg]
\qquad \nonumber \\
&+&(-\gamma^R\otimes\tilde{\Delta}^R-\Sigma^R)\otimes x 
+x \otimes (-\Delta^A\otimes\tilde{\gamma}^A+\Sigma^A)
\nonumber \\
&=&
-\gamma^R\otimes\tilde{\Sigma}^K\otimes\tilde{\gamma}^A+\Delta^K\otimes
\tilde{\gamma}^A+\gamma^R\otimes\tilde{\Delta}^K-\Sigma^K. \nonumber \\
\end{eqnarray}
Summation over the repeated index $j$ is implied.  The expressions for
$\tilde{f}^{R}$, $f^A$, and $\tilde{g}^K$ are obtained by comparing
Eq.~(\ref{gl_green3}) with Eqs.~(\ref{ricc1})--(\ref{ricc3}).

\section{Results: Diffusive Case}

As an example we study for this case the heterostructure
shown on the right side of Fig. \ref{eps1.1}, which contains
a weak diffusive ferromagnet between two superconductors.
A domain wall of Bloch type is centered in the ferromagnetic part.
Problems involving domain walls have previously been
treated only with techniques that do not allow access to information
on spatial variations, \cite{cht} or elaborate methods involving
rotating coordinate systems.\cite{ber} However, since the
Riccati description contains the full 2$\times$2 spin structure, it
extends easily to account for any such situation.

We adopt for the description of a weak ferromagnetic materials ($J$ 
much smaller than the Fermi energy)
a quasiparticle dispersion given by a spin-dependent energy shift
$
E\hat\tau_3 \rightarrow E\hat\tau_3-\vec{J}({\bf R}) \cdot \hat{\vec{\sigma}},
$
in Eq.~(\ref{gl_usdl}). Here $\vec{J}({\bf R})$ denotes the effective
exchange field of the ferromagnet, and ${\vec{\sigma}}$ denotes
the vector of Pauli spin matrices. 
We solve the Usadel equations using the Riccati parameterization
for a ferromagnetic material, which contains a domain wall of the
Bloch type. We model the domain wall by a $\vec{J}$-vector rotating across the domain wall with  the
polar angle $\theta_J= \mbox{atan} (z/d_W)$, where $d_W$ parameterizes the width
of the domain wall.

\begin{figure}
\sidecaption
\includegraphics[width=0.6\textwidth]{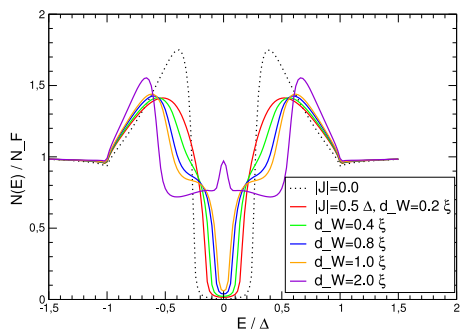}
\caption[]{Local density of states as a function of energy in the center of a
heterostructure containing a weak ferromagnet ($J=0.5 \Delta $) of
length $d_F=2\xi_0$ between two superconductors. A
Bloch domain wall of length $d_W$ is centered in the ferromagnet
(see Fig. \ref{eps1.1}). Results for various $d_W$ are shown}
\label{eps5.1}
\end{figure}

In Fig. \ref{eps5.1} we present results for the local density of states
at the center position of the heterostructure shown in Fig \ref{eps1.1}.
The length of the ferromagnet is $d_F=2\xi_0$, where $\xi_0=\sqrt{D/\Delta }$
is the coherence length in the ferromagnet, and $\Delta $ is the gap
in the superconductors. The dotted curve shows the behavior for
a normal metal between two superconductors, showing the well known
minigap. The effect of a domain wall in the center of the ferromagnet
is shown for several wall thicknesses $d_W$ in Fig. \ref{eps5.1}.
The new structures in the presence of the domain wall in Fig. \ref{eps5.1}
are due to spin-triplet correlations which mix between all 
three spin-triplet channels due to the continuous variation of the spin
quantization axis across the domain wall \cite{ber}. This leads to the appearence of additional
Andreev bound states inside the minigap, modifying the total density of states.
In general, the presence of the domain wall reduces the minigap,
and for thick enough domain walls the minigap closes completely.
This suggests that the width of the domain walls can
influence the transport properties of such devices considerably.

\section{Conclusions}
We have developed a framework for studying heterostructures with superconducting
and ferromagnetic parts, and have applied it to two cases: a ferromagnet 
in the ballistic regime with strong spin polarization ($J\sim E_F$); and
a weak ferromagnet ($J\ll E_F$) in the diffusive regime, having an
inhomogeneous, non-collinear spin magnetization (Bloch domain wall).
In both cases we have studied the superconducting proximity effect with 
the ferromagnet sandwiched between two superconductors.
We have shown, that singlet-triplet mixing occurs near interfaces 
between superconductors and strong ferromagnets, or near domain-wall structures 
in the case of weak ferromagnets in proximity to a superconductor.
The origin of equal-spin triplet correlations in the two studied cases is very different. 
In a strong ferromagnet between two superconductors, 
the singlet triplet mixing takes place within the superconductor, in
a layer of the order of the coherence length near the interface,
enabling triplet correlations to penetrate into the ferromagnet.  
In contrast, in weak ferromagnets with domain walls of
Bloch type, equal-spin triplet correlations
are produced within the ferromagnet, near the  domain walls.
\section{Acknowledgements}
This work was supported by Deutsche Forschungsgemeinschaft within the
Center for Functional Nanostructures.

%

\end{document}